\documentstyle[12pt]{article}
 at 14.4truept at 12.0truept
\title{ The Sigma model formulation of the  type II string theory in the $Ads_5\times
S_5$ backgrounds with Ramond-Ramond Flux}
\author{
{\large Satchidananda  Naik} \thanks{e-mail: naik@mri.ernet.in}
\\
  Harish-chandra Research Institute \\
 Chhatnag Road, Jhusi  \\
Allahabad-211 019, INDIA\\}

\begin{document}
\maketitle

\hspace*{\fill}

\hspace*{\fill}
\newcommand{\bee}{\begin{equation}}
\newcommand{\nn}{\nonumber}
\newcommand{\ee}{\end{equation}}
\newcommand{\ba}{\begin{array}}
\newcommand{\ea}{\end{array}}
\newcommand{\bea}{\begin{eqnarray}}
\newcommand{\eea}{\end{eqnarray}}
\newcommand{\ki}{\chi}
\newcommand{\eps}{\epsilon}
\newcommand{\pa}{\partial}
\newcommand{\lb}{\lbrack}
\newcommand{\Se}{S_{\rm eff}}
\newcommand{\rb}{\rbrack}
\newcommand{\de}{\delta}
\newcommand{\th}{\theta}
\newcommand{\rh}{\rho}
\newcommand{\ka}{\kappa}
\newcommand{\al}{\alpha}
\newcommand{\bt}{\beta}
\newcommand{\si}{\sigma}
\newcommand{\bsi}{\Sigma}
\newcommand{\vp}{\varphi}
\newcommand{\gm}{\gamma}
\newcommand{\gb}{\Gamma}
\newcommand{\om}{\omega}
\newcommand{\et}{\eta}
\newcommand{\gt}{ {g^2 T }\over{4 {\pi}^2}}
\newcommand{\qab}{{{\sum}_{a\neq b}}{{q_a q_b}\over{R_{ab}}}}
\newcommand{\omb}{\Omega}
\newcommand{\pr}{\prime}
\newcommand{\ra}{\rightarrow}
\newcommand{\nb}{\nabla}
\newcommand{\MSb}{{\overline {\rm MS}}}
\newcommand{\lnh}{\ln(h^2/\Lambda^2)}
\newcommand{\cz}{{\cal Z}}
\newcommand{\h}{{1\over2}}
\newcommand{\Lm}{\Lambda}
\newcommand{\inft}{\infty}
\newcommand{\bpa}{\bar {\partial}}  
\newcommand{\hth}{\hat {\theta}}
\newcommand{\hp}{\hat p}
\newcommand{\hb}{\hat b}
\newcommand{\hc}{\hat c}
\newcommand{\hbt}{\hat {\beta}}
\newcommand{\hgm}{\hat {\gamma}}
\newcommand{\hvp}{\hat {\varphi}}
\newcommand{\hlm}{\hat {\lambda}}    
\newcommand{\hv}{\hat v} 
\newcommand{\hq}{\hat q} 
\newcommand{\Lra}{\Longleftrightarrow}
\newcommand{\abschnitt}[1]{\par \noindent {\large {\bf {#1}}} \par}
\newcommand{\subabschnitt}[1]{\par \noindent
                                          {\normalsize {\it {#1}}} \par}
%-----------------------------------------------------------------------
% The definition below makes spaces e.g \skipp{3} makes 3 spaces
\newcommand{\skipp}[1]{\mbox{\hspace{#1 ex}}}
 
%
%
% various slashed symbols
%
%
%\newcommand\slash#1{\rlap{$#1$}/} % slashes a character
\newcommand\dsl{\,\raise.15ex\hbox{/}\mkern-13.5mu D}
    % this one can be subscripted
\newcommand\delsl{\raise.15ex\hbox{/}\kern-.57em\partial}
\newcommand\Ksl{\hbox{/\kern-.6000em\rm K}}
\newcommand\Asl{\hbox{/\kern-.6500em \rm A}}
\newcommand\Dsl{\hbox{/\kern-.6000em\rm D}} %roman D
\newcommand\Qsl{\hbox{/\kern-.6000em\rm Q}}
\newcommand\gradsl{\hbox{/\kern-.6500em$\nabla$}}
%--------
 %---------------------------------------------------------------
 %\vskip5.0cm
\newpage
\begin{abstract} \normalsize
A sigma model action is constructed for the type II string in the $Ads_5\times S_5$
back grounds with Ramond-Ramond flux.
\end{abstract}

\vskip10.0cm
   
\newpage
\pagestyle{plain}
\setcounter{page}{1}
\abschnitt{1. Introduction}

The conjecture of duality between D=4 and $N=4$ super Yang- Mills theory and the type
IIB string theory compactified on a five dimensional anti-de Sitter space and a 
five sphere $Ads_5\times S_5$  has drawn global attention in recent past
 \cite{Mald,Pol,Witt}.
It is not very straight forward to quantize super string with the  $Ads_5\times S_5$
background geometry since there exists self-dual Ramond-Ramond 5-form flux in the
Neveu-Schwarz-Ramond (NSR) formulation of string theory. Several attempts are made to
quantize string theory in the  $Ads_5\times S_5$ back ground. The Green-Schwarz 
formulation  of string theory in this background is consistently studied classically
\cite{Tsey} and quantization of this is attempted \cite{Raj}. 
%%%%%%%%%%%%%%%%%%%%%%%%%%%%%%%%%%%%%%%%%%%%%%%%%%%%%%%%%%%%%%%%%%%
%%%%%%%%%%%%%%%%%%%%%%%%%%%%%%%%%%%%%%%%%%%%%%%%%%%%%%%%%%%%%%%%%%%%%%%%%%%%%%%%%%
One faces several technical problems when attempts to covariantly quantize the 
superstring even in the flat ten dimensional background. Only with the Neveu-Schwarz
background, the world-sheet super symmetry is intact so BRST structure and the
Picture-changing  operations are smooth for the calculation of physical states. 
However the Ramond sectors are always represented by the Spin fields. They break the
world sheet supersymmetry and give rise to ambiguities of Picture-changing. The
spin-spin field  correlations give rise to square root cuts and space-time Super
symmetry becomes obscure. To illustrate,  
the space-time supersymmetry generator in the $-~\h $ picture is given by
 \bee 
q_{\al} = \int dz e^{-{\frac{1}{2}\vp}} S_{\al}
\ee                                                                       
where $\vp$ comes from the fermionization of the Bosonic ghosts which is usually
 defined as $\beta =\pa \xi e^{-\vp}$, $\gamma = \eta e^\vp$ and $S^{\al}$ 's are the
spin fields of conformal weight $5/8$ constructed from
the $\psi^{\mu}$, the Neveue-Schwarz vector. The supersymmetry (SUSY) algebra is
closed on-shell
\bee
 \{ q^-_{\al}, q^-_{\bt}\} =  {\gb }^{\al\bt}_{\mu}
\oint dz e^{-{\frac{1}{2}\vp}}\psi^{\mu}
\ee
where $e^{-{\frac{1}{2}\vp}}\psi^{\mu}$ is not the standard momentum operator. By
using a Picture changing operator $P~=~\{Q, \xi \}$ where $Q$ is the BRST operator,
one gets
\bee
P\oint dz e^{-{\frac{1}{2}\vp}}\psi^{\mu} = \oint dz \pa x_{\mu}
\ee
the standard momentum operator. Thus one defines one of the generators in the
$+\h $ picture and another in the  $-\h $ picture to get the explicit  SUSY
\bee
\{ q^+_{\al}, q^-_{\bt}\} =  {\gb }^{\al\bt}_{\mu} \oint dz \pa x_{\mu}
\ee

algebra. However it has double the  number of required $N=1$ ten dimensional  
SUSY generators and it is essential to maintain  both the ten dimensional
Poincare invariance and the correct number of SUSY generators. Recently
Berenstein and Leigh \cite{BL} have an alternative approach to derive sigma
model action in the $Ads_5\times S_5$ background. They take the Ramond operator
as $P_{\h}e^{-{\frac{1}{2}\vp}} S_{\al} $ where $P_{\h}$ is the holomorphic square
root of the picture changing operator $P$. Then perturbing 
with Ramond-Ramond vertex operators, around the flat background                            
 they are able to
generate the $Ads_5\times S_5$
background by Fischler-Susskind mechanism\cite{FS}. The conformal invariance is shown
up to two loops.

         In the recent past Berkovits and his collaborators \cite{Berk1, Berk2,
Berk3, Berk4} chose
a novel way to tackle this problem. They redefine the Neveue-Schwarz-Ramond (NSR)
variables like Green- Schwarz type of variables such as ${\th}^{\al}~= ~
e^{\frac{1}{2}\vp} S^{\al}$ and its canonical conjugate
 $p_{\al} = e^{-{\frac{1}{2}\vp}}S_{\al} $ . However ${\th}^{\al}$ variables are not
all free fields for $\al = 1,..., 16$. There exists OPE among themselves as 
\bee
{\th}^{\al} (y) {\th}^{\bt}( z) \to {\frac{1}{(y-z)}} {\gb }^{\al\bt}_{\mu}
e^{\vp}\psi^{\mu}.
 \ee
 Here $ {\gb }^{\al\bt}_{\mu}$ are the ${\bf 16\times 16}$ Pauli matrices in 10
dimensions.  It is 
quite   possible to choose a subset of these variables which will be free
fields. This can happen when four or six of the dimensions are compactified and one
has explicitly $SO(6)$ or $SO(4)$ invariance instead of $SO(9,1)$ invariance.
In Ref.\cite{Berk2} Berkovits et al. have picked four of the free $\th $ variables
and their canonical conjugates to show the explicit SUSY invariance in six dimensions.
Then  $R^6$ is deformed to $Ads_3\times S_3$
by adding  vertex operator for the   Ramond-Ramond three form in the action. The
Ramond-Ramond form breaks the SO(6) symmetry to $ SO(3)\otimes SO(3)$ which is
locally isomorphic to $SO(4)$  and  is the  remnant of the bosonic symmetry.
Thus the Lagrangian describes a supersymmetric sigma model in the group manifold
of $SU(2)\times SU(2)$ which can be identified with  $Ads_3\times S_3$  space.
The same procedure cannot be extended to ten dimension since it is not possible
to find eight of the free fermionic coordinates to describe manifestly ten
dimensional supersymmetry. Maximally one can have five free fermionic coordinates
   so one  can have $U(5)$ super Poincare invariance \cite{Berk5}.

To avoid these problems we have a new formulation of  fermionic coordinates
 which are  linear combination of $S_a$ and $S^{\dot a}$ so that
  we can  keep $SO(8)\times SO(2)$ type of invariance.      
 Thus  $p_{\al}$  and ${\th}_{\al}$ in terms of $8_S$ and $8_C$ of $SO(8)$ are
\bea
 p_a =& ({\si}^1 S_a + C_{ a{\dot a}} {\si}^2 S^{\dot a}) \h e^{-{\frac{1}{2}\vp}}
\\
 p_{\dot a} =& ({\si}^1 S_{\dot a} - C_{ a{\dot a}} {\si}^2 S^{\dot a}) \h
e^{-{\frac{1}{2}\vp}} \\
 {\th}^a =& ({\si}^1 S^{ a} + C^{ a{\dot a}} {\si}^2 S_{\dot
a}) \h e^{{\frac{1}{2}\vp}} \\ 
{\th}^{\dot a }=& ({\si}^1 S^{\dot a} - C^{ a{\dot a}}
{\si}^2 S_{\dot a}) \h e^{{\frac{1}{2}\vp}}
\eea      
where ${\si}^1$ ,  ${\si}^2$ and  ${\si}^3$  are Pauli matrices.  
Here  $ C^{ a{\dot a}}$ is symmetric and somewhat like  the charge conjugation matrix 
of $SO(6)$ Clifford algebra where $ C {\gm}^t_{i} C^{-1}~ = ~ - {\gm}_i $. The
commutation relations and the operator product expansions (OPE) are given as  
\bea
 p^{ij}_a ( z) {\th}^b_{jk}(w) &=& - {\frac{1}{2 (z - w)}} {\de }^i_k {\de }^a_b \\
p^{ij}_{\dot a }( z) {\th}^{\dot b}_{jk}(w) &=&
 - {\frac{1}{2 (z - w)}} {\de }^i_k {\de }^{\dot a}_{\dot b}
\eea              
 the other OPE s are 
\bee
 p_a p_b = {\th}^a {\th}^b = p_{\dot a} p_{\dot b} = {\th}^{\dot a} {\th}^{\dot b} = 0
\ee
 and also 
\bee 
p_a{\th}^{\dot b} = p_{\dot a}{\th}^b =0.
 \ee 
Here $p_{\dot a}$ and
${\th}_{\dot a}$ 
are not independent variables since
 \bee
 p_a = -i {\si}^3 C_{ a{\dot a}} {\th}^{\dot a}e^{- \vp} .
 \ee                                                     
This gives  8 $p_a$ and 8 ${\th}^a$ 
which are independent variables. This has the correct number of $N=1$ susy
fermionic variables. We exactly follow here the procedure of Berkovits et al.
\cite{Berk2} to derive sigma model action with  $Ads_5\times S_5$ as  back ground
from the flat ten dimensional superspace by perturbing with Ramond-Ramond
vertex operator. Our fermionic coordinates are derived from the SUSY generators
with $ - \h$ picture. We express the SUSY generators with $+\h$ picture with these
fermionic  coordinates only and hence we have to couple to the ghosts. Thus we have
 only $N=1$ SUSY in the target space where as we have $N=2$ SUSY in the world-sheet.  
In this limitations we cannot write an action with arbitrarily  curved background.
However this can be done taking another set of superspace variables with $+ \h $
picture and use the harmonic constraints.

In section 2. we describe the world-sheet and the space-time supersymmetry
of our action. In section 3. we show how to deform $R^{10}$ to $Ads_5\times S_5$
by adding the vertex operator of the Ramond-Ramond five form in the action. We show
that the remnant bosonic symmetry is $SU(4)\times SU(4)$ which indicates that the
target space to be $SU(4|4)$ super group manifold. In section 4. we conclude
very briefly and the gamma matrices are explicitly given in the appendix. 

\newpage
\abschnitt{2. The Space time Supersymmetry formulation}
 In the $-\h $ picture we
define our supersymmetric generator as
 \bee
 q^-_{\al} = \oint dz j^-_{\al} = \int dz
p_{\al} 
\ee
 where
 \bee
 p_{\al} = \pmatrix{p_a \cr p_{\dot a}\cr}
 \ee 
>From the OPEs
(c.f. (7)--(12)) we can show that
 \bee 
Tr\left( \{ q^-_{\al}, q^-_{\bt}\} \right)= 0 ,
 \ee 
where
$Tr$ is over the $\si$ matrices. Applying the picture changing operator $\{ Q,\xi \}$
 on   $j^-_{\al}$ we get 
\bee 
j^+_{\al} = b \eta e^{ 2 \vp} j^-_{\al} + {\bar j}^{\al}_{+}
\ee    

where ${\bar j}^{\al}_{+} = ({\bar j}^a_{+}~~~~~ {\bar j}^{\dot a}_{+} )$ 
and
 \bee
{\bar j}^a_{+ ij} = \left({\gb}^{\mu}_{ab} {\th}^b_{ij} + {\gb}^{\mu}_{a{\dot b}}
{\th}^{\dot b}_{ij}\right){\pa}x_{\mu}
 \ee
 and 
\bee 
{\bar j}^{\dot a}_{+ ij} = \left({\gb}^{\mu}_{{\dot a}b} {\th}^b_{ij} +
{\gb}^{\mu}_{{\dot a}{\dot b}}{\th}^{\dot b}_{ij}\right) {\pa}x_{\mu}.
\ee
This gives OPE
\bee
\{ j^{- ij}_{\al}(z) , {\bar j}^{\bt}_{+ jk}(w)\} ~
= ~ {\frac{1}{2 (z -w)}}{\de }^i_k {\gb}^{\mu}_{\al \bt} {\pa}x_{\mu}
\ee
which is  the correct covariant off-shell space time supersymmetric commutation
relation.

The action is
\bea
S &=& {\frac{1}{2\pi}}\int \Big [{\bpa }x_{\mu}{\pa} x^{\mu} + S_{ghost} 
 ~ + p_{\al}{\bpa}{\th}^{\al} + {\hp}_{\al}\pa{\hth}^{\al}\nn\\
&& + {\lambda}^a \left(p_a - i {\si}^3  C^{ a{\dot a}}e^{- \vp}{\th}^{\dot a} \right) 
 + {\lambda}^{\dot a }\left (p_{\dot a } +  i {\si}^3  C^{ a{\dot a}}e^{- \vp}{\th}^a\right)\nn 
\\
&& +  {\hlm}^a \left(p_a - i {\si}^3 C^{ a{\dot a}}e^{- \hvp}{\hth}^{\dot a} \right)
 + {\hlm}^{\dot a }\left(p_{\dot a } +  i {\si}^3 C^{ a{\dot a}}e^{- \hvp}{\hth}^ a\right)\Big] \nn\\
\eea
where
\bee
S_{ghost} = S_{bc} + S_{\bt \gm} + S_{ge}.
\ee
Here $S_{ge}$ denotes the extra non-interacting
ghosts to make total central charge zero.
The right moving fields are all denoted with a hat. Here ${\lambda}^{\al}$'s are all
Lagrange multiplier to use the constraint of eq.(14 ). 
 We use here
all the ghosts in the bosonized form as
$b = e^{- i \si}$ and $c = e^{i \si}$ and $\eta = e^{ i \ka}$ and $\xi = e^{- i \ka}$.

The OPE's of free fields are
\bee
x^{\mu}(z) x^{\nu}(w) = {\eta}^{\mu \nu} \log (z -w)^2,~~ 
 {\ka}(z){\ka}(w) = ~~ {\si}(z){\si}(w) = - \log(z -w)
\ee

The energy momentum tensor is
\bee
T = \h \pa x_{\mu}\pa x^{\mu} + p_{\al}\pa {\th}^{\al} +T_{ghost}.
\ee

To express  the super Virasore generator we need the expression of $\psi$ in terms of
the superspace variable $\th$. Thus we first express spin fields as
\bee
 S^a = (1 + i) \Big[ {\si}^1 {\th}^a e^{-\h \vp} +
 C^{a{\dot a}} {\si}^2 p_{\dot a}e^{\h \vp}\Big] 
\ee
 and
 \bee
 S^{\dot a}= (1 - i) \Big[{\si}^1 {\th}^{\dot a}e^{-\h \vp} +
 C^{ a{\dot a}} p_a  {\si}^2 e^{\h \vp}\Big].                 
\ee
We define here
\bea
A^a &= & e^{\h \vp}S^a\nn\\
 B^a &= & e^{{-{\frac{3}{2}} } \vp}S^a\nn\\
 C^{\dot a} &= & e^{\h \vp}S^{\dot a} \nn\\
D^{\dot a} &= &  e^{{-{\frac{3}{2}} } \vp}s^{\dot a} 
\eea
so that  we can denote $\psi$ as
\bea
\psi_{+} &= & {\gb}_{ab} A^a B^b e^{\vp}\nn\\
\psi_{-} &= & {\gb}_{{\dot a}{\dot b}} C^{\dot a} D^{\dot b }e^{\vp}\nn\\
\psi_{i} &= & {\gb}_{{\dot a}b} A^{\dot a} C^b e^{\vp}.
\eea
The  super Virasore generator for the critical $N = 1 $ system is
\bee  
G = {\psi}_{\mu} \pa x^{\mu}.
\ee
which can be expressed in terms of new world sheet variables.
The BRST operator is
\bee
Q_{BRST} = {\frac{1}{2\pi}}\oint dz \left[ cT + i b\pa c + \h \pa \vp \pa \vp 
\\ + {\pa}^2 \vp +i \eta \pa \xi + i \eta e^{\vp} G + i b\eta \pa \eta e^{2\vp} \right]
\ee

We can have a twisted $N=2$ super Virasore algebra by taking $G^{+} = J_{BRST}$ and
$G^{-} = b$.

\newpage
\abschnitt{3. The Sigma model  formulation}
   
We deform the flat ten dimensional space-time to $Ads_5\times S^5$ by adding a
vertex operator $V_H$  depicting Ramond-Ramond flux.
Let's denote a self dual five form as $i H_{01234} = H_{56789}$. The vertex operator
due to this will break $SO(10)$ symmetry to $SO(5)\times SO(5)$ rotational symmetry.
We set  this to be a constant
\bee
i H_{01234} = H_{56789} = 2 g
\ee
where $g$ is a small parameter measuring the strength of the coupling.
This is taken to be a constant so that it can  maintain the translational symmetry of 
the $R^{10}$. The vertex operator in the $-\h $ picture is $V^{--}_H = q^-{\hat q}^- $
and in the $+ \h $   picture $V^{++}_H = q^+{\hat q}^+ $. We take here the most
general
one as the linear combination of both of them.
\bea
V_H  &=&  g \Big[ q^-_{\al}{\hat q}^-_{\al} + q^{+}_{\al}{\hat q}^{+}_{\al} \Big] \nn\\
     &=& g\int \left[ p_{\al} {\hp}_{\al} + \left( p_{\al} e^{\phi} - {\gb}^{\mu}_{\al \bt}
{\th}^{\bt} \pa x_{\mu} \right)\left( {\hp}_{\al} e^{{\hat \phi}} - {\gb}^{\mu}_{\al
\bt}{\hth}^{\bt} {\bpa}x_{\mu} \right)\right] \eea 
where  $\phi =i(\ka -\si ) + 2 \vp $. 
After  a scaling $p_{\al} = g^{ -\h}p_{\al}$, $\th = g^{ -\h}\th$ and 
$e^{\phi} = - g e^{\phi}$ one gets 
\bee 
  V_H  =  \int \Big[ p_{\al} {\hp}_{\al}  -
 g^2  \left( p_{\al}e^{\phi} + {\gb}^{\mu}_{\al \bt} {\th}^{\bt} \pa x_{\mu} \right)\left(
{\hp}_{\al} e^{{\hat \phi}} +  {\gb}^{\mu}_{\al \bt}{\hth}^{\bt} {\bpa}x_{\mu} \right) \Big]    
\ee
In order to get a sigma model type of action  we rescale all the fields \\
$(x, p_{\al}, {\th}^{\al}, e^{\phi},{\hp}_{\al},{\hth}^{\al},.)$ to 
$g^{-1} (x, p_{\al}, {\th}^{\al}, e^{\phi},{\hp}_{\al},{\hth}^{\al},.)$  to get
\bea
 S &=&  g^{-2}\int \Big[ \pa x{\bpa}x + p_{\al}{\bpa}{\th}^{\al} +
{\hp}_{\al}\pa{\hth}^{\al}\nn\\ 
&& + p_{\al} {\hp}_{\al} +{\lambda}^{\al}p_{\al} +
{\hlm}^{\al}{\hp}_{\al}\nn\\ 
&& + \left( p_{\al}e^{\phi} + {\gb}^{\mu}_{\al \bt} {\th}^{\bt}
\pa x_{\mu} \right) \left({\hp}_{\al} e^{{\hat \phi}} + {\gb}^{\mu}_{\al \bt}{\hth}^{\bt}
{\bpa}x_{\mu} \right)\nn \\ 
&& + i {\si}^3  C^{ a{\dot a}} \left\{ e^{- \vp}\left({\lambda}^{\dot a} {\th}^a -
{\lambda}^a{\th}^{\dot a}\right) + e^{\hvp} \left( {\hlm}^{\dot a}{\hth}^a - {\hlm}^a
{\hth}^{\dot a} \right)\right \}\Big]
\eea

Here $e^{\phi}$ and $e^{{\hat \phi}}$  are zero weight conformal fields.
To start with we  set them zero  and separately consider them
perturbatively  since they scale like the perturbative parameter g.         
Here  $p_{\al}$ and ${\hp}_{\al}$ are like auxiliary variables so we integrate 
these variables by using their equation of motion which gives

\bee 
{\bpa}{\th}^{\al} +  {\lambda}^{\al} = - {\hp}_{\al}
\ee
and 
\bee
\pa{\hth}^{\al} + {\hlm}^{\al} = - p_{\al}.
\ee

 Substituting these we get
\bea
S &=&  g^{-2}\int \Big[ \pa x{\bpa}x + {\lambda}^{\al}{\hlm}^{\al} -
 {\bpa}{\th}^{\al}\pa{\hth}^{\al}\nn\\
&&+  {\gb}^{\mu}_{\al \bt}{\gb}^{\nu}_{\al \gm}{\th}^{\bt}{\hth}^{\gm}\pa
x_{\mu}{\bpa}x_{\nu}\nn\\
&&+ i {\si}^3  C^{ a{\dot a}} \left \{ e^{- \vp}\left({\lambda}^{\dot a} {\th}^a 
-{\lambda}^a{\th}^{\dot a}\right) + e^{\hvp} \left( {\hlm}^{\dot a}{\hth}^a - {\hlm}^a  
{\hth}^{\dot a} \right)\right \}\Big]
\eea
 
The most general  supersymmetric transformations of various fields generated by
$v^{\al}_{\pm}q^{\pm}_{\al} + {\hv}^{\al}_{\pm}{\hq}^{\pm}_{\al}$ are
\bee
\de {\th}^{\al} = v^{\al}_{-}  +  v^{\al}_{+}e^{\phi} +
 i {\si}^3 v^{\bt}_{+}{\tilde F}_{\bt \al} + {\hv}^{+}_{\bt}
({\gb}^{\mu})^{\al \bt}x_{\mu}
\ee
and
\bee
\de x^{\mu} = v^{\bt}_{+}({\gb}^{\mu})_{\bt \al}{\th}^{\al} +
{\hv}^{+}_{\bt}({\gb}^{\mu})_{\bt \al}{\hth}^{\al}
\ee
where
\bee
{\tilde F}_{\bt \al} = \left[({\gb}^{\mu})_{\bt a}C^{a\dot a}{\de}_{{\dot a}\al} +
({\gb}^{\mu})_{\bt{\dot a}}C^{a\dot a}{\de}_{a \al} \right]F_{\mu}
\ee
and
\bee
F_{\mu} = \int \frac{e^{\vp}}{(z - w)^{\frac{3}{2}}} \pa x_{\mu}
\nn\ee

If we drop the term of $e^{\phi}$ and $v^{-}_{\al}$ which is like the translation in
the superspace we observe that   $x,\th , {\hat{\th}}$ transform under susy on to
themselves and we get a rotation in the super space. 
To verify the number of bosonic generators  we take
\bee
Tr\{ {\de}^2 x_{\mu}\} =  v^{\al}_{+}\left[ ({\gb}^{\mu})_{\al \bt}
{\tilde{\gb}}^{t \bt \gm }_{\nu}  - {\tilde{\gb}}^{ \al \bt}_{\mu}
({\gb}^{t \nu})_{\bt \gm } \right]{\hv}^{\gm}_{+} x_{\nu}   \\
= R_{\mu \nu} x_{\nu}
\ee
Here $R_{\mu \nu}$ is the pure rotation of the bosonic coordinate $x_{\mu}$ which
has 30 independent parameters.{\footnote{ In the $SO(8)$ decompositions there are
three
antisymmetric $\gm$ and five symmetric $\gm$ matrices (c.f. Appendix).Thus rotational
parameters are
\bea
& v^a (\gm_i -{\gm}^t_i ){\hv}^{\dot a}\nn \\
&v^{\dot a}(\gm_i -{\gm}^t_i ){\hv}^a\nn
 \eea give 6 parameters. Similarly 
\bea
 &v^a ( \gm_i{\gm}^t_j -{\gm}^t_j \gm_i){\hv}^a
\nn\\ 
&v^{\dot a} ( \gm_i{\gm}^t_j -{\gm}^t_j \gm_i){\hv}^{\dot a}\nn 
\eea 
give 18
parameters and 
\bea
 &v^a ( \gm_i{\gm}^t_8 - {\gm}^t_i \gm_8){\hv}^a \nn\\
&v^{\dot a} ( \gm_i{\gm}^t_8 - {\gm}^t_i\gm_8){\hv}^{\dot a}\nn 
\eea
give 6 more elements. This gives total 30 matrix elements. }}

Now if we integrate over $\lambda$ and $\hlm$ and neglecting the terms
which couples to b$\eta$ ghosts we get
\bee
S =  g^{-2}\int \Big[ \pa x{\bpa}x - {\bpa}{\th}^{\al}\pa{\hth}^{\al}
+  {\gb}^{\mu}_{\al \bt}{\gb}^{\nu}_{\al
\gm}{\th}^{\bt}{\hth}^{\gm}\pa x_{\mu}{\bpa}x_{\nu}
+ e^{\vp + \hvp}{\th}^{\al}{\gb}^9{\hth}^{\al} + x^{\mu} x^{\nu}\pa x_\mu{\bpa}x_{\nu}\Big]
\ee

Now we treat 
\bee
{\th}^a_q = \pmatrix{\th^a \cr \hth^{\dot a}\cr \th ^{\dot a}\cr\hth^{ a}\cr } 
\ee
as a quartet of the continuous symmetry of $SU(4)$ which is clearly  seen  if take the
second
variation of $\th^{\al}$ due to supersymmetry. Besides here is   a symmetry
 for    $z\Lra {\bar z}$
\bee
{\th}^a_q = {\epsilon}_{qp}{\th}^a_p
\ee
where ${\epsilon}_{qp} $ is an antisymmetric ${\bf 4\times 4}$  matrix $i
{\si}^1\times
{\si}^2$.
So without the $b\eta$ ghost coupling
we get the action
\bee
S =  g^{-2}\int \Big[ {\pa}_i x^{\mu}{\pa}_i x_{\mu} - {\epsilon}_{qp} {\pa}_i{\th}^a_p
{\pa}_i{\th}^a_q + x^{\mu} x^{\nu}{\pa}_i x_\mu{\pa}_ix_{\nu} 
+ e^{\vp + \hvp}({\th}^a_1 {\th}^a_4 - {\th}^a_3 {\th}^a_2 )\Big]
\ee
The last term is due to the constraint that originally ${\th}^a$ and ${\th}^{\dot a}$
are not all independent and there exists OPE among them. That is if  explicitly
calculated it  halves the 8 components of ${\th}^a$ so 
that  we  have total 16
components of fermions instead of the present one with 32 components. This action is
a perfect sigma model action with a Lagrangian of the form of $G_{IJ}{\pa}^i{\Phi}^I
{\pa}_i{\Phi}^J$ where $G_{IJ}$ is the metric of the $SU(4|4)$ and $\Phi$ are the 
coordinates on the group manifold.

 \abschnitt{4. Conclusion} 
In this article we present the preliminary version of our big endeavour and there are
many more things which deserve further investigations. Our construction of fermionic
variables enable us to make the theory covariant. We get clearly two sets of fermionic
variables those are non-interacting. We never face the complications of quantizing
the theory with local fermionic symmetry with mixed first class and second class
constraints which usually plagues the  covariant quantization problem. The coupling of
zero weight ghost field $e^{\phi}$ and $e^{\tilde \phi}$ is neglected here in order to
make the action look simple. Our  next step is to take this in to account.
The bosonic rotations of $x_{\mu}$ has 30 elements (c.f. eqn.(43)) and the four
copies of   fermionic coordinates (c.f.eqn (42)) indicate that the target space is a
homogeneous space of $SU(4|4)$ super group manifolds. To verify our model we can make
a  semi classical analysis as has been done by  de Boer et al. \cite{ooguri} for the
$Ads_3$ and compare  the correlation functions of $<T(x)T(y)>_{world sheet}$
with that of the target space on the boundary. There is no problem in finding the
dilaton -dilaton correlation function and compare with the results of Maldacena
conjecture.
We are not repeating these things here. The whole purpose of our exercise is to
extend this things to higher genus Rieman surface and take all the ghost couplings
which we keep it for the further investigation.

 \abschnitt{5. Appendix}

\begin{eqnarray} 
&& \Gamma^0=\left(\begin{array}{cc} {\bf 1}_8 & 0\\ 0 & {\bf1}_8\end{array}\right),\;\;
 \tilde\Gamma^0=\left(\begin{array}{cc} -{\bf 1}_8 & 0\\ 0 
& -{\bf 1}_8\end{array}\right),\cr && \Gamma^i=\left(\begin{array}{cc} 0 &
{\gamma^i}_{a\dot a}\\ 
\bar\gamma^i{}_{\dot aa} & 0 \end{array}\right), \;\;
 \tilde\Gamma^i=\left(\begin{array}{cc} 0 & {\gamma^i}_{a\dot a}\\
\bar\gamma^i{}_{\dot aa} & 0\end{array}\right),\cr &&
\Gamma^9=\left(\begin{array}{cc} {\bf 1}_8 & 0\\ 0 & -{\bf 1}_8\end{array}\right),
\;\;
 \tilde\Gamma^9=\left(\begin{array}{cc} {\bf 1}_8 & 0\\ 0
 & -{\bf 1}_8\end{array}\right).
\end{eqnarray}                    

Here ${\gamma^i}_{a\dot a}$, $\bar\gamma^i{}_{\dot aa}\equiv -
({\gamma^i}_{a\dot a})^{\rm T}$ are  $SO(8)$ $\gamma$-matrices 
for i = 1,...7 and 
 ${\gamma^8} = {\bf 1}_8 $  which obey 
 \begin{equation}
\gamma^i\bar\gamma^j+\gamma^j\bar\gamma^i=2\delta^{ij}{\bf 1}_8, 
\end{equation}
and $i,a,\dot a=1,\dots,8$. 
More explicitly
\bea
\gm_1 &= & - i {\si}^2\times {\si}^2 \times {\si}^2  \nn\\
\gm_2 &= &i {\si}^2\times {\si}^3 \times {\bf 1}  \nn\\ 
\gm_3 &= & i{\si}^2\times {\si}^1 \times {\bf 1}   \nn\\ 
\gm_4 &= & - i{\si}^2\times {\si}^2 \times {\si}^1   \nn\\ 
\gm_5 &= & i {\si}^1\times {\bf 1} \times {\bf 1}    \nn\\ 
\gm_6 &= &i {\si}^3\times {\bf 1}  \times {\bf 1}   \nn\\
\gm7  &=& -i{\si}^2\times {\si}^2 \times {\si}^3      \nn\\
 \gm_8 &= & {\bf 1}\times {\bf 1}\times {\bf 1}         \nn\\
\eea
Similarly

\bee
C = {\si}^2\times {\bf 1}\times {\si}^2.
\nn
\ee
   
 \end{document}